\documentclass{PoS}

\usepackage{graphicx}

\newcommand{\note}[1]{}                    

\newcommand{\mnote}[1]{}                   

\newcommand{\sixth}{\mbox{\small $\frac{1}{6}$}}         
\newcommand{\third}{\mbox{\small $\frac{1}{3}$}}         

\def\lsim{\mathrel{\rlap{\lower4pt\hbox{\hskip1pt$\sim$}}
    \raise1pt\hbox{$<$}}}                
\def\gsim{\mathrel{\rlap{\lower4pt\hbox{\hskip1pt$\sim$}}
    \raise1pt\hbox{$>$}}}                

\title{
\vspace*{-1.25cm}
\begin{minipage}{\textwidth}
\begin{flushright}
\texttt{\footnotesize
PoS(LATTICE2014)110  \\
ADP-14-40/T899       \\
DESY 14-231          \\
Edinburgh 2014/21    \\
Liverpool LTH 1029   \\
}
\end{flushright}
\end{minipage}\\[15pt]
\vspace*{+1.25cm}
       Determining Sigma - Lambda mixing}

\ShortTitle{$\Sigma$ -- $\Lambda$ mixing}

\author{\speaker{R. Horsley}$^{\,a}$,
        J. Najjar$^b$,
        Y. Nakamura$^c$,
        H. Perlt$^d$,
        D. Pleiter$^{eb}$, 
        P.~E.~L. Rakow$^f$,
        G. Schierholz$^g$,
        A. Schiller$^d$,
        H. St\"uben$^h$
        and J.~M. Zanotti$^i$ \\
        \llap{$^a$} School of Physics and Astronomy,
                    University of Edinburgh,
                    Edinburgh  EH9 3FD, UK \\
        \llap{$^b$} Institut f\"ur Theoretische Physik,
                    Universit\"at Regensburg, 93040 Regensburg, Germany \\
        \llap{$^c$} RIKEN Advanced Institute for Computational Science,
                    Kobe, Hyogo 650-0047, Japan \\
        \llap{$^d$} Institut f\"ur Theoretische Physik,
                    Universit\"at Leipzig, 04109 Leipzig, Germany \\
        \llap{$^e$} JSC, Forschungszentrum J\"ulich,
                    52425 J\"ulich, Germany \\
        \llap{$^f$} Theoretical Physics Division,
                    Department of Mathematical Sciences,
                    University of Liverpool,
                    Liverpool L69 3BX, UK \\
        \llap{$^g$} Deutsches Elektronen-Synchrotron DESY,
                    22603 Hamburg, Germany \\
        \llap{$^h$} Regionales Rechenzentrum, Universit\"at Hamburg,
                    20146 Hamburg, Germany \\
        \llap{$^i$} CSSM, School of Chemistry and Physics,
                    University of Adelaide, Adelaide SA 5005, Australia \\
        E-mail: \email{rhorsley@ph.ed.ac.uk} }

\author{QCDSF-UKQCD Collaborations}

\abstract{SU2 isospin breaking effects in baryon octet (and decuplet) masses
          are due to a combination of up and down quark mass differences
          and electromagnetic effects. These mass differences are small.
          Between the Sigma and Lambda the mass splitting is much larger,
          but this is mostly due to their different wavefunctions.
          However there is now also mixing between these states.
          We determine the QCD mixing matrix and hence find the
          mixing angle and mass splitting.}

\FullConference{The 32nd International Symposium on Lattice Field Theory,\\
		23-28 June, 2014\\
		Columbia University New York, NY}

\begin{document}


\section{Introduction}

\begin{figure}[h]
   \begin{center}
      \includegraphics[width=4.50cm]{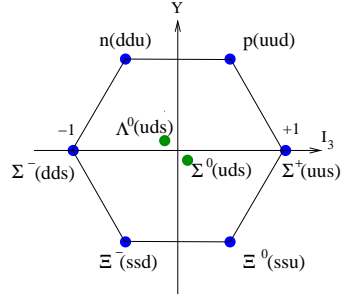}
   \end{center}
\caption{The baryon octet in the $I_3$--$Y$ plane.}
\label{baryon_octet}
\end{figure}
\begin{figure}[h]
\begin{center}

\begin{minipage}{0.45\textwidth}

   \begin{center}
      \includegraphics[width=7.00cm]{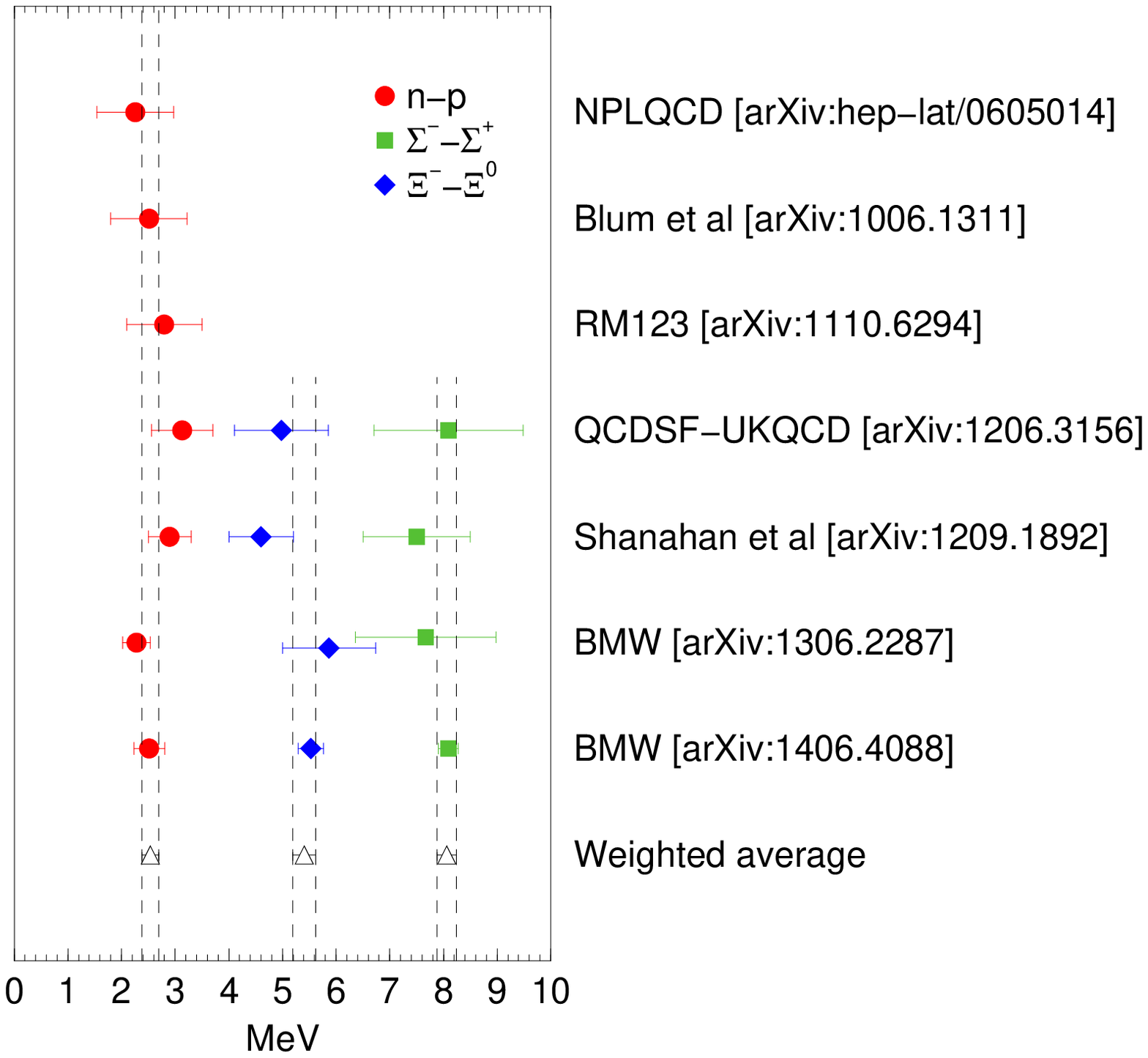}
   \end{center} 

\end{minipage}\hspace*{0.05\textwidth}
\begin{minipage}{0.45\textwidth}

   \begin{center}
      \includegraphics[width=7.25cm]{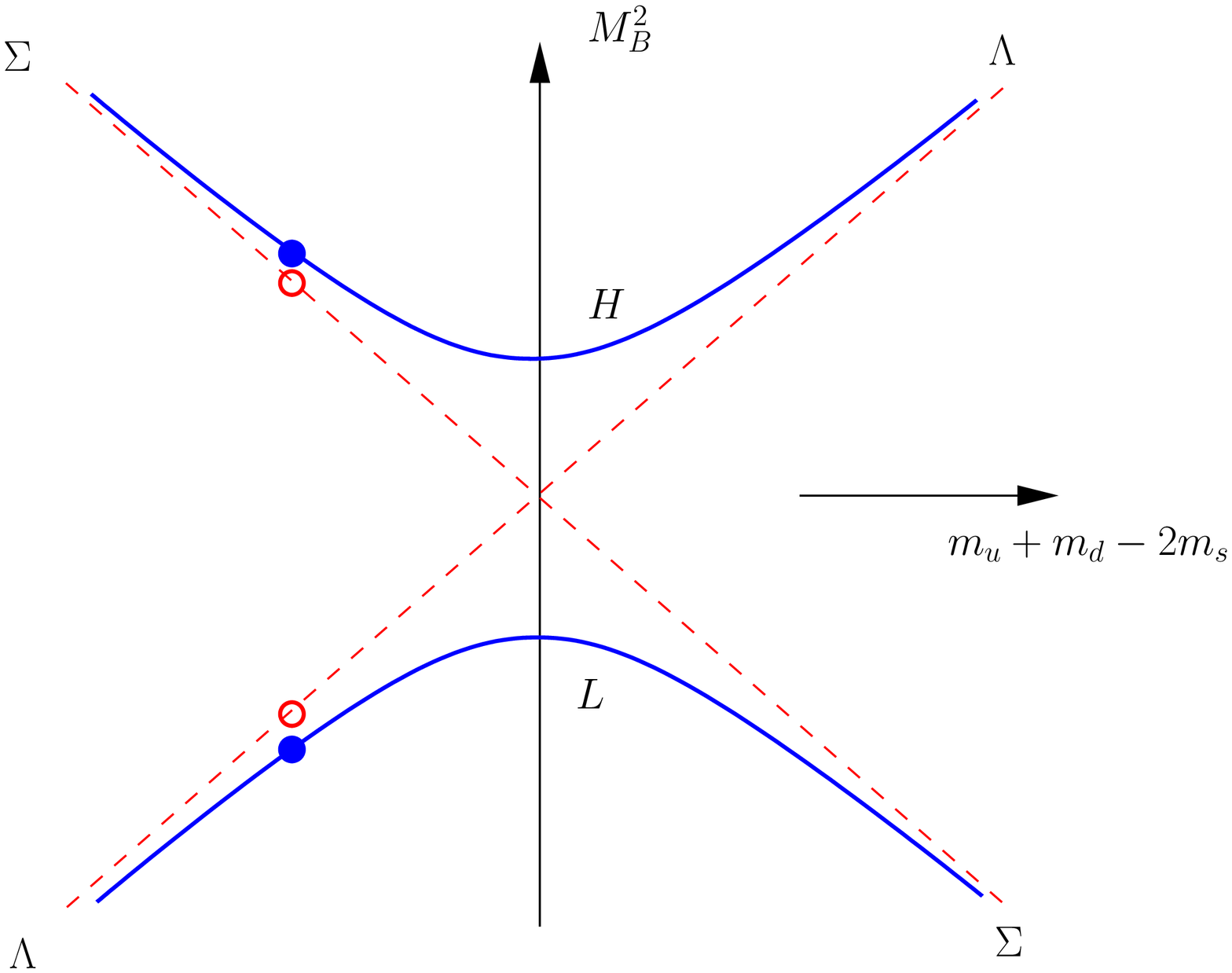}
   \end{center} 

\end{minipage}
\caption{Left panel: Lattice determinations for isospin mass breaking
         due to $u$--$d$ quark mass differences for
         $n$--$p$, $\Sigma^-$--$\Sigma^+$ and $\Xi^+$--$\Xi^0$,
         together with a weighted average.
         Right panel: A sketch of the heavy, $H$, and light, $L$,
         baryon $(\mbox{masses})^2$ against $m_u+m_d-2m_s$ for fixed
         $m_u - m_d$. The mass splitting between the Sigma and
         Lambda masses in the isospin limit ($m_u = m_d$) is given
         by the difference between the (red) dashed lines;
         if $m_u \not= m_d$ then there is an additional mass
         difference due to mixing, as given by the (blue) lines. 
         The physical point is indicated by the filled (blue) circles.}
\label{mass_split+mixing}

\end{center}
\end{figure}

$SU2$ isospin breaking effects in hadron octet (and decuplets)
are due to a combination of up and down quark mass differences
and electromagnetic effects%
\footnote{QED effects will not be considered here.}.
The baryon octet is shown in the $I_3$--$Y$ plane in Fig.~\ref{baryon_octet}.
On the baryon octet `outer' ring the effects of $u$--$d$ mass differences
are very small $\sim O(\mbox{few MeV})$. (The difference
in masses between the $Y = \mbox{const}.$ particles in this figure.)
A compilation of some lattice determinations of these mass splittings
is given in the left panel of Fig.~\ref{mass_split+mixing}.
However for the Sigma and Lambda baryons, sitting at the centre
of the octet, the mass splitting is much larger,
$(M_{\Sigma^0} - M_{\Lambda^0})^{\exp} = 76.959(23) \,\mbox{MeV}$.
This is mainly due to their different wavefunctions.
However despite the fact that both particles have the same
quark content ($u$, $d$, $s$) there is also a small additional
isospin component due to mixing between these states when the $u$ and
$d$ quarks have different masses, as depicted in the right panel
of Fig.~\ref{mass_split+mixing}. We have the situation of
`avoided level crossing'. All lines are at constant $m_u - m_d$,
the (red) dashed lines are for the isospin limit $m_u = m_d$,
while the (blue) lines are for the case  $m_u - m_d \not= 0$.
The centre point is when all quark masses are the same.
We denote the two branches by `$H$' and `$L$'.
The mass splitting between the Sigma and Lambda particles
is given by the vertical difference between these points.

In this talk we determine the $\Sigma$--$\Lambda$ mixing matrix
and hence find the mixing angle and mass splitting.
Further details and results are given in \cite{horsley14a}.


\section{Method}


The strategy we employ here has been described in
\cite{bietenholz11a,horsley12a}; we shall extend it here to cover
the mixing case. Briefly, in lattice simulations (and in particular for
the case considered here of three flavours) there are many paths for
the quark masses to approach the physical point. We have chosen here
to extrapolate from a point on the $SU(3)$ flavour symmetry line
(when all the quark masses are equal to $m_0$ say) to the physical point.
As will shortly be seen it is sufficient to consider this for the
case of degenerate $u$ and $d$ quark masses (i.e.\ $m_u = m_d \equiv m_l$,
together with the strange quark mass $m_s$). Thus we take
$(m_0, m_0) \rightarrow(m^*_l, m^*_s)$ (where a $*$ denotes the
physical point). To define the path the choice here is to
keep the singlet quark mass $\overline{m}$ constant, where
$\overline{m} = m_0 = \third ( 2m_l + m_s)$, along the trajectory.
We now develop the $SU(3)$ flavour symmetry breaking Taylor expansion
for hadron masses beginning at the flavour symmetric point in
terms of
\begin{eqnarray}
   \delta m_q  = m_q - \overline{m} \,.
\label{deltamq}
\end{eqnarray}
The expansion coefficients are functions of $\overline{m}$ alone and
the path is called the `unitary line' as we expand in both sea and
valence quarks (with the same masses). Thus provided $\overline{m}$
is kept constant, then the expansion coefficients in the Taylor
expansion remain unaltered whether we consider $2+1$
or $1+1+1$ flavours (i.e.\ mass degenerate $u$ and $d$ quark masses
or not). This opens the possibility of determining quantities
that depend on $1+1+1$ flavours from just $2+1$ flavour simulations.

Furthermore we can generalise the $SU(3)$ flavour breaking expansion
to the case of partially quenched (PQ) valence quark masses, $\mu_q$
(with possibly different masses to the sea quark masses $m_q$)
without increasing the number of expansion coefficients%
\footnote{The advantage of using PQ valence quarks is that they are
          computationally cheaper.}
. Equivalently to eq.~(\ref{deltamq}) we set
\begin{eqnarray}
   \delta \mu_q  = \mu_q - \overline{m} \,.
\end{eqnarray}

We now define a quark mass matrix ${\cal M}$ and baryon mass matrix
$M({\cal M})$ where
\begin{small}
\begin{eqnarray}
   {\cal M} = \left( \begin{array}{ccc}
                        m_u  & 0  & 0   \\
                        0   & m_d & 0   \\
                        0   & 0   & m_s \\
                     \end{array}
              \right) \,, \quad
   M^2({\cal M}) 
   &=& \pmatrix{
       M^2_n & 0 & 0 & 0 & 0  & 0 & 0 & 0 \cr  
       0 & M^2_p  & 0 & 0 & 0  & 0 & 0 & 0 \cr  
       0 & 0 & M^2_{\Sigma^-}  & 0 & 0 & 0  & 0 & 0 \cr  
       0 & 0 & 0 & M^2_{\Sigma\Sigma} & M^2_{\Sigma\Lambda} & 0 & 0 & 0 \cr  
       0 & 0 & 0& M^2_{\Lambda\Sigma} & M^2_{\Lambda\Lambda} & 0 & 0 & 0\cr
       0 & 0 & 0 & 0 & 0 & M^2_{\Sigma^+}  & 0 & 0 \cr  
       0 & 0 & 0 & 0 & 0 & 0 & M^2_{\Xi^-}  & 0 \cr  
       0 & 0 & 0 & 0 & 0 & 0& 0  & M^2_{\Xi^0}    
            } \,,
\label{matrices}
\end{eqnarray}
\end{small} 
and demand%
\footnote{The $SU(3)$ flavour breaking expansion holds for any function
of the baryon mass matrix; we have found that using $M_B^2$ gives
(slightly) better fits than $M_B$ alone.}
that under all $SU(3)$ transformations
\begin{eqnarray}
   {\cal M} \to {\cal M^\prime} = U{\cal M}U^\dagger 
   \quad \leftrightarrow \quad
   M^2({\cal M}^\prime) = UM^2({\cal M})U^\dagger \,.
\end{eqnarray} 
Mathematically under these transformations there is no
change to the eigenvalues; physically there is also no change,
possibly just a relabelling (e.g.\ $m _d \leftrightarrow m_s$ is
equivalent to relabelling $ M_n \leftrightarrow M_{\Xi^0}, \ldots$).
We write $M^2 = \sum_{i=1}^{10} K_i(m_q,\mu_q) N_i$,
where the $N_i$ matrices are classified under $S_3$ and $SU(3)$ symmetry
and the $K(m_q,\mu_q)$ are coefficients. The $S_3$ symmetry group is that
of the (equilateral triangle $C_{3v}$) and has $3$ irreducible
representations: two singlets $A_1$, $A_2$ and  one doublet $E$
with elements $E^\pm$. The $N_i$ are mostly diagonal, e.g.\
$N_1 = \mbox{diag}(1,1,1,1,1,1,1,1)$, except $N_5$, $N_8$, $N_{10}$,
where the $\Sigma$ -- $\Lambda$ $2\times 2$ sub-matrices are
non-diagonal. Further details of the diagonal matrices are given in
\cite{bietenholz11a}; the complete set is described in \cite{horsley14a}.

This gives for baryons, $B(aab)$ with valence quarks $a, b, c$
on the outer ring of the octet
\begin{eqnarray}
   M^2_{B} &=& P_{A_1} + P_{E^+} \,,
\label{Baab}
\end{eqnarray}
and for the baryons $B(abc)$ at the centre of the octet (i.e.\ the
$2\times 2$ submatrix in $M^2$ in eq.~(\ref{matrices}))
\begin{eqnarray}
   \left( \begin{array}{cc}
             M_{\Sigma \Sigma}^2 & M_{\Sigma \Lambda}^2 \\
             M_{\Lambda\Sigma}^2 & M_{\Lambda\Lambda}^2 \\
          \end{array} \right)
     =  P_{A_1} \left( \begin{array}{cc}
                           1 & 0 \\
                           0 & 1 \\
                        \end{array} \right)
        + P_{E^+} \left( \begin{array}{cc}
                           1 & 0 \\
                           0 & -1 \\
                        \end{array} \right)
        + P_{E^-} \left( \begin{array}{cc}
                           0 & 1 \\
                           1 & 0 \\
                        \end{array} \right)
        + P_{A_2} \left( \begin{array}{cc}
                           0 & -i \\
                           i & 0 \\
                        \end{array}
                \right) \,.
\label{2x2submatrix}
\end{eqnarray}
The $P_G$ are functions of the quark masses with the 
symmetry $G$ under the $S_3$ permutation group and are given
to NLO as
\begin{eqnarray} 
   P_{A_1} 
      &=& M_0^2 + 3A_1 \delta\overline{\mu}
                                                          \nonumber   \\
      & & + {\textstyle{1\over 6}} B_0 
                      ( \delta m_u^2 + \delta m_d^2 + \delta m_s^2)  
          + B_1 ( \delta\mu_a^2 + \delta\mu_b^2  + \delta\mu_c^2 ) 
                                                          \nonumber   \\
      & &      + {\textstyle{1\over 4}} (B_3 + B_4)     
          \left[ (\delta\mu_c - \delta\mu_a)^2 
                + (\delta\mu_c - \delta\mu_b)^2 + (\delta\mu_a - \delta\mu_b)^2 
          \right] + O(3)
                                                          \nonumber   \\
   P_{E^+} 
      &=& {\textstyle{3\over 2}} A_2 ( \delta\mu_c - \delta\overline{\mu} )
                                                          \nonumber   \\
      & & + {\textstyle{1\over 2}} B_2 
                           ( 2 \delta\mu_c^2 - \delta\mu_a^2 - \delta\mu_b^2)
                                                          \nonumber   \\
      & & + {\textstyle{1\over 4}} (B_3 - B_4)  
          \left[ (\delta\mu_c - \delta\mu_a)^2 
                + (\delta\mu_c - \delta\mu_b)^2 -2 (\delta\mu_a - \delta\mu_b)^2
          \right] + O(3)
                                                          \nonumber   \\
   P_{E^-} 
      &=& {\textstyle{\sqrt{3}\over 2}} A_2 (\delta\mu_b - \delta\mu_a) 
                                                          \nonumber   \\
      & & + {\textstyle{\sqrt{3}\over 2}} B_2 (\delta\mu_b^2 - \delta\mu_a^2)
          + {\textstyle{\sqrt{3}\over 4}} (B_3 - B_4) 
            \left[(\delta\mu_c - \delta\mu_b)^2 - (\delta\mu_c - \delta\mu_a)^2
            \right] + O(3)
                                                          \nonumber   \\
    P_{A_2}
        &=& 0 + O(3) \,,
\end{eqnarray}
($\delta\overline{\mu} = (\delta\mu_a+\delta\mu_b+\delta\mu_c)/3$).
NNLO (i.e.\ $O(3)$) terms have also been determined, \cite{horsley14a}. 
Diagonalisation of eq.~(\ref{2x2submatrix}) yields
\begin{eqnarray}
   M^2_H = P_{A_1} + \sqrt{P_{E^+}^2 + P_{E^-}^2 + P_{A_2}^2}\,, \qquad
   M^2_L &=& P_{A_1} - \sqrt{P_{E^+}^2 + P_{E^-}^2 + P_{A_2}^2} \,.
\label{M2H+M2L}
\end{eqnarray}
Although looking rather complicated, in the isospin limit when there
is no mixing, these expansions reduce to those given in
\cite{bietenholz11a}. Writing the eigenvectors as
$e_H = ( \cos\theta, e^{-i \phi} \sin\theta )$ and
$e_L = ( -e^{i \phi} \sin\theta, \cos\theta )$
gives for the mixing angle $\theta$, and phase, $\phi$ 
\begin{eqnarray}
   \tan 2\theta = {\sqrt{P_{E^-}^2 + P_{A_2}^2} \over P_{E^+}} \,,
   \qquad 
   \tan\phi = {P_{A_2} \over P_{E^-}} \,,
\end{eqnarray}
and close to the physical point we set
$M_{\Sigma^0} = M_H$, $M_{\Lambda^0} = M_L$ (and
$\theta_{\Sigma^0\Lambda^0} = \theta$).

Practically, when analysing the raw lattice results for
the baryon masses, it is better to use scale invariant ratios 
(which helps to make the data smoother).
We define the scale implicitly using singlet quantities
$X_S$, $S = \pi, N, \ldots\,$. For the octet baryons it is
convenient to define a `centre of mass' quantity
\begin{eqnarray}
   X_N^2 &=& \sixth( M_p^2 + M_n^2 + M_{\Sigma^+}^2 +  M_{\Sigma^-}^2
                                + M_{\Xi^0}^2 + M_{\Xi^-}^2 ) 
                                                                \nonumber \\
        &=& M_0^2 + \sixth(B_0+B_1+B_3)
                    (\delta m_u^2 + \delta m_d^2 + \delta m_s^2) + O(3) \,.
\end{eqnarray}
Experimentally $X^{\exp}_N = 1.160\,\mbox{GeV}$. All singlet quantities
have no $O(\delta m_q)$ terms and we have seen \cite{bietenholz11a}
that they remain constant down to the physical point,
enabling a reliable determination of the scale.
It is convenient to form dimensionless ratios (within a multiplet)
\begin{eqnarray}
  \tilde{M}^2 \equiv {M^2 \over X^2_S} \,, \quad 
            S = \pi, N, \ldots \,,
            \qquad
  \tilde{A}_i \equiv {A_i \over M_0^2} \,,\quad
  \tilde{B}_i \equiv {B_i \over M_0^2} \,, 
         \end{eqnarray}
and use this in the Taylor expansions.

For example this gives for $\Sigma$ -- $\Lambda$ mixing
at LO in the unitary limit, the analytic results
\begin{eqnarray}
   \tilde{M}_{\Sigma^0} - \tilde{M}_{\Lambda^0}
      = \sqrt {3 \over 2} 
           \tilde{A}_2 \sqrt{\delta m_u^2 + \delta m_d^2 + \delta m_s^2}\,,
   \qquad
   \tan 2\theta = {(\delta m_d - \delta m_u) \over \sqrt{3}\delta m_s}  \,.
\label{M_siglam_diff}
\end{eqnarray}
This shows clearly that any mass difference is dominated by the $\tilde{A}_2$
coefficient as the $\tilde{A}_1$ terms have cancelled. This is different
to the baryons on the outer ring, which are a mixture of the $\tilde{A}_1$
and $\tilde{A}_2$ coefficients (and the numerical values mean that
it is actually dominated by the $\tilde{A}_1$ coefficient). Note also
that in the isospin limit (where there is no mixing), the mass
square root in eq.~(\ref{M_siglam_diff}) simplifies considerably
to give $\sqrt{6}\delta m_l$.


\section{Results}


We use here an $O(a)$ NP improved clover action with tree level Symanzik
glue and mildly stout smeared $2+1$ clover fermions, \cite{cundy09a},
at $\beta = 5.50$ on $32^3\times 64$ and $48^3\times 96$ sized lattices.
We have found that $\kappa_0 = 0.12090$ provides a suitable starting
point on the $SU(3)$ symmetric line. The quark mass (whether valence
or unitary) is defined as $\mu_q = (1/\kappa_q-1/\kappa_{0c})/2$,
where $\kappa_{0c}$ is the critical $\kappa_0$ in the chiral limit
along the $SU(3)$ symmetric line. However this does not need to be
determined as in $\delta\mu_q$ it cancels.

\begin{figure}[h]

\begin{minipage}{0.45\textwidth}

   \begin{center}
      \includegraphics[width=6.50cm]
                   {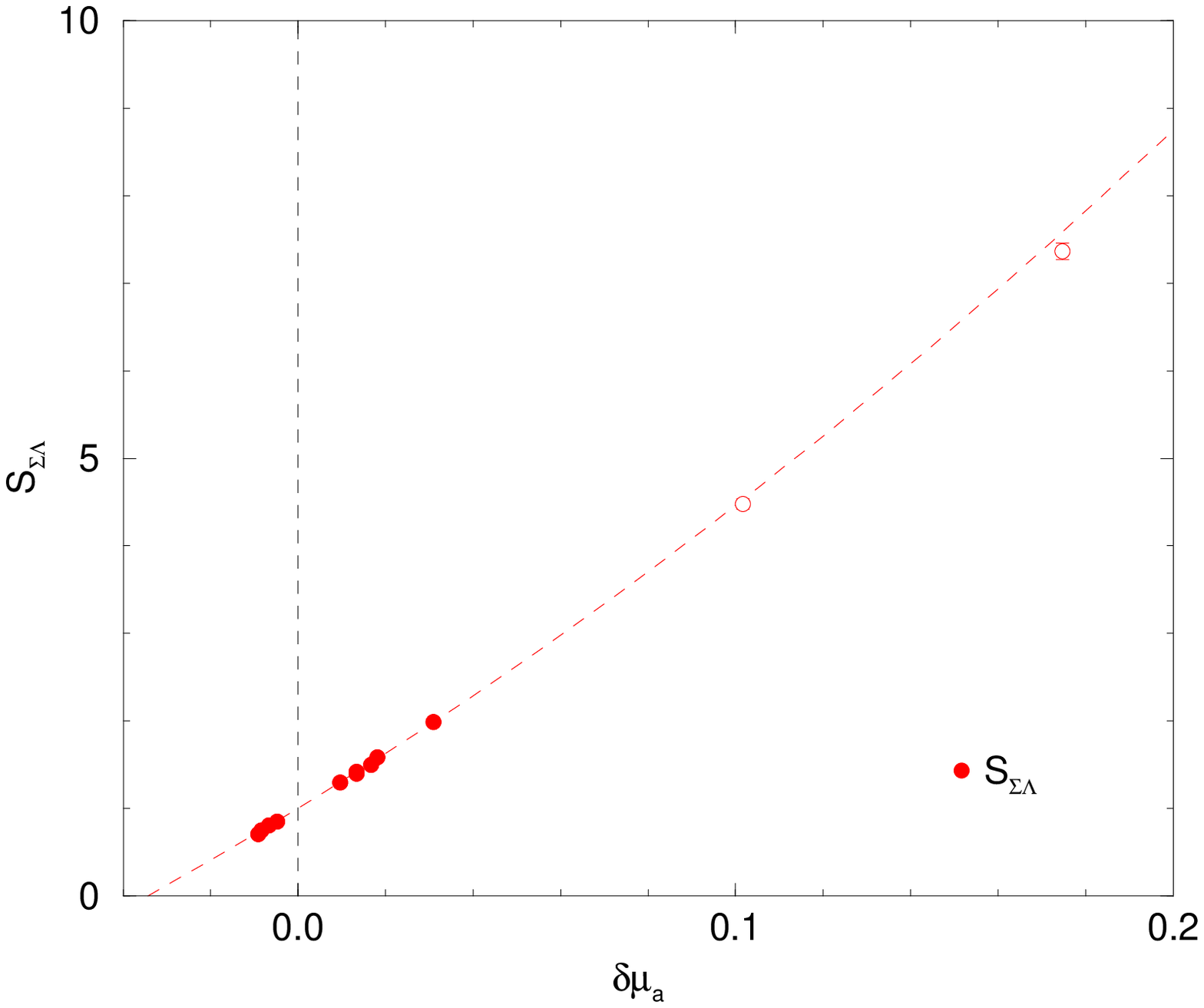}
   \end{center} 

\end{minipage}\hspace*{0.05\textwidth}
\begin{minipage}{0.45\textwidth}

   \begin{center}
      \includegraphics[width=7.00cm]
          {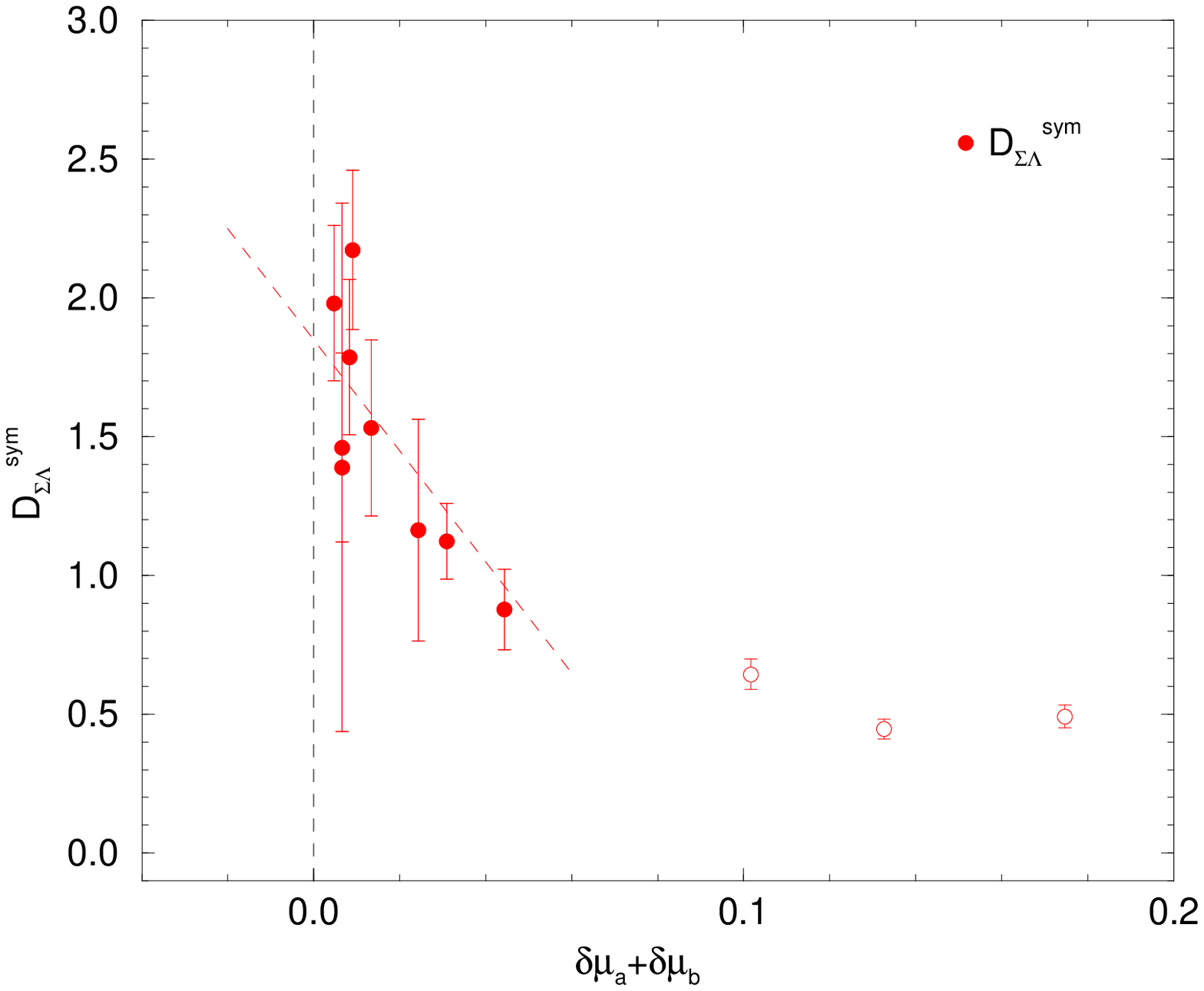}
   \end{center} 

\end{minipage}
\caption{Left panel: $S_{\Sigma\Lambda}$ from eq.~(\protect\ref{S}).
         Right panel: $D_{\Sigma\Lambda}^{\rm sym}$ from
         eq.~(\protect\ref{Dsym}). Both are plotted against
         $\delta\mu_a + \delta\mu_b$. Points used in the fit
         are denoted by filled circles.}
\label{baryon_S+D}
\end{figure}

The method is first to determine the physical quark masses using
the pion octet and equivalent expansions to those described above
(and of course only considering pseudoscalar particles on the 
outer ring), by fitting to unitary and PQ data. This is described
in \cite{horsley12a} and we also use the results from there.
We then for the baryon octet use the unitary and PQ data to determine
the $\tilde{A}$ and $\tilde{B}$ coefficients. To be sure that the
$SU(3)$ flavour expansion is valid we restrict quark masses to a
range here taken to be $|\delta\mu_a|+|\delta\mu_b|+|\delta\mu_c| \lsim 0.2$.
This translates to nucleon masses of $\lsim 2\, \mbox{GeV}$.
(In fits it was then found that $\tilde{B}_3$ was then compatible with zero.)
Two simple plots which illustrate the situation are the completely
mass degenerate case (when $\Sigma$ and $\Lambda$ are the same)
\begin{eqnarray}
   S_{\Sigma\Lambda}
     \equiv \tilde{M}_\Sigma^2(aaa^{\prime\prime})
     = 1 + 3\tilde{A}_1\delta\mu_a + 3\tilde{B}_1\delta\mu_a^2 \,,
\label{S}
\end{eqnarray}
($a^{\prime\prime}$ is a mass degenerate but distinct quark) and the
`symmetric' difference case (between $\Sigma$ and $\Lambda$)
\begin{eqnarray}
   D^{\rm sym}_{\Sigma\Lambda}
      &\equiv& { \tilde{M}_\Sigma^2(aab) - \tilde{M}_\Lambda^2(aa^\prime b) 
          - \tilde{M}_\Sigma^2(bba) + \tilde{M}_\Lambda^2(bb^\prime a)
          \over
          4(\delta\mu_b - \delta\mu_a) }
      = \tilde{A}_2 + \tilde{B}_2(\delta\mu_a+\delta\mu_b) \,,
\label{Dsym}
\end{eqnarray}
as shown in Fig.~\ref{baryon_S+D}.
For $S_{\Sigma\Lambda}$, the fit is very good and could be easily
extended. As mentioned before $\tilde{A}_1$ is the relevant
coefficient for mass splittings on the outer baryon ring.
For $D_{\Sigma\Lambda}^{\rm sym}$ the symmetric difference
is chosen in order to minimise possible effects of
terms involving  $\delta\mu_a - \delta\mu_b$.
The plot has a sharp increase as the quark mass is reduced,
and presumably a non-polynomial behaviour there. As this is related
to the $\Sigma$--$\Lambda$ mass splitting, this necessitates
a restricted fit region. (It should be noted that the
unitary quark masses have $|\delta m_a| \lsim 0.01$.)
The reason for this behaviour is due to spin--spin interaction
between the quarks. From the Dirac equation we expect the magnetic
moment to be $\propto 1/m_a$, which might suggest a spin--spin
interaction of the form $\propto 1/(m_am_b)$. This has
also recently been proposed in \cite{yang14a}.

\begin{figure}[htb]
   \begin{center}
      \includegraphics[width=8.00cm]
  {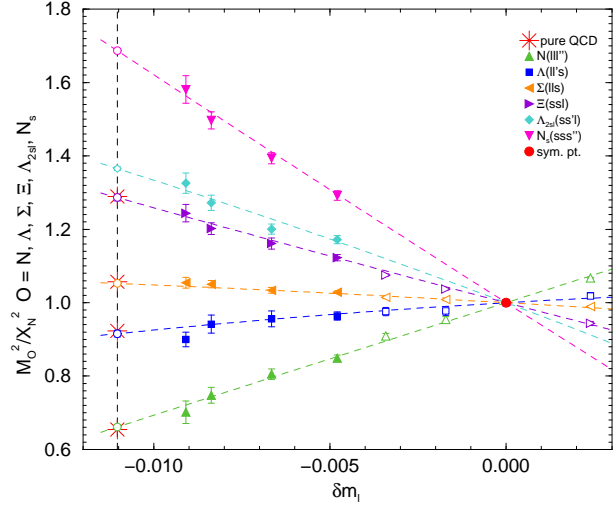}
   \end{center}
   \vspace*{-0.25in}
\caption{The baryon `fan' plot for the `$N$' and `$\Lambda$' type particles
         $\tilde{M}_{N_O}^2$ ($N_O = N$, $\Sigma$, $\Xi$, $N_s$) and
         $\tilde{M}_{\Lambda_O}^2$ ($\Lambda_O = N$,
         $\Lambda$, $\Lambda_{l2s}$, $N_s$) versus $\delta m_l$.
         The symbols are all unitary data.
         (The opaque triangular symbols are from comparison
         $24^3\times 48$ sized lattices and not used in the fits here.)
         The common symmetric point is the filled circle.
         The vertical dashed line is the $N_f = 2+1$ pure QCD physical point,
         with the opaque circles being the numerically determined pure
         QCD hadron mass ratios for $2+1$ quark flavours.
         For comparison, the stars represent
         the average of the $(\mbox{mass})^2$ of
         $M_N^{*\,2}(lll^{\prime\prime}) = (M_n^{\exp\,2}(ddu) + M_p^{\exp\,2}(uud))/2$,
         $M_\Lambda^{*\,2}(lls) = M_{\Lambda^0}^{\exp\,2}(uds)$,
         $M_\Sigma^{*\,2}(lls)
                 = (M_{\Sigma^-}^{\exp\,2}(dds) + M_{\Sigma^+}^{\exp\,2}(uus))/2$
         and
         $M_\Xi^{*\,2}(ssl) 
                 = (M_{\Xi^-}^{\exp\,2}(ssd) + M_{\Xi^0}^{\exp\,2}(ssu))/2$.}
\label{b5p50_dml_mNmLamaaboXN2-jnt_N+L}
\end{figure}

Secondly we show a `fan' plot for the $2+1$ flavour case:
$\tilde{M}_N^2(aab)$, $\tilde{M}_\Lambda^2(aa^\prime b)$,
in Fig.~\ref{b5p50_dml_mNmLamaaboXN2-jnt_N+L}.
We have $N(lll^{\prime\prime}) [= \Lambda_{3l}(ll^\prime l^{\prime\prime})]$,
$\Sigma(lls)$, $\Xi(ssl)$,
$N_s(sss^{\prime\prime}) [= \Lambda_{3s}(ss^\prime s^{\prime\prime}]$
and $\Lambda(ll^\prime s)$, $\Lambda_{l2s}(ss^\prime l)$.
($N_s(sss^{\prime\prime})$ and $\Lambda_{l2s}(ss^\prime l)$ are fictitious
baryons, but provide additional useful data for the fits.)
As this is the diagonal case there is no mixing and from
eqs.~(\ref{Baab}), (\ref{M2H+M2L}) $\tilde{M}_N^2 = P_{A_1} + P_{E^+}$,
$\tilde{M}_\Lambda^2 = P_{A_1} - P_{E^+}$. We find good agreement
with the expected `physical' results.

For baryons on the outer ring of the octet we find that the
central values of the mass splittings are in good agreement with
previous results, \cite{horsley12a} (see also the left panel
of Fig.~\ref{mass_split+mixing}), however with an increased error bar.
This is the result of the situation
depicted in Fig.~\ref{baryon_S+D} where previously
as shown in the left panel plot, we were able to use
a larger fit range. For $\Sigma^0$ and $\Lambda^0$ we find
\begin{eqnarray}
   M_{\Sigma^0} - M_{\Lambda^0} =  79.44(7.37)(3.37) \,\mbox{MeV}\,, \qquad
   \tan 2\theta_{\Sigma^0\Lambda^0} = 0.0123(45)(25) \,.
\end{eqnarray}
As anticipated, this gives a very small $\theta_{\Sigma^0\Lambda^0}\lsim 1^\circ$.
Taking the difference between the $M_{\Sigma^0} - M_{\Lambda^0}$
and $M_\Sigma^*(lls)-M_\Lambda^*(lls)$ gives the contribution due to
isospin breaking of $\sim 0.01\,\mbox{MeV}$.


\section*{Acknowledgements}


The numerical configuration generation (using the BQCD lattice
QCD program) and data analysis 
(using the Chroma software library) was carried out
on the IBM BlueGene/Q using DIRAC 2 resources (EPCC, Edinburgh, UK),
the BlueGene/P and Q at NIC (J\"ulich, Germany), the
SGI ICE 8200 and Cray XC30 at HLRN (Berlin--Hannover, Germany)
and on the NCI National Facility in Canberra, Australia
(supported by the Australian Commonwealth Government).
This investigation has been supported partly 
by the EU grants 227431 (Hadron Physics2) and 283826 (Hadron Physics3).
JN was partially supported by EU grant 228398 (HPC-EUROPA2).
HP was supported by DFG Grant: SCHI 422/9-1.
JMZ was supported by the Australian Research
Council grants FT100100005 and DP140103067.
We thank all funding agencies.



\end{document}